\documentclass[12pt]{article}

\usepackage{amsmath,amssymb,amsthm}

\newcommand{\real}{\mathbb{R}}

\newcommand{\vj}{\boldsymbol{j}}
\newcommand{\vv}{\boldsymbol{v}}

\newcommand{\vs}{\boldsymbol{s}}

\newcommand{\vtri}{\boldsymbol{\vartriangle}}

\newcommand{\Gammav}{\varGamma}

\begin{document}

\title{\bf  
Zero-Energy Flows and Vortex Patterns 
\break
in Quantum Mechanics 
}

\author{Tsunehiro Kobayashi\footnote{E-mail: 
kobayash@a.tsukuba-tech.ac.jp} \\
{\footnotesize\it Department of General Education 
for the Hearing Impaired,}
{\footnotesize\it Tsukuba College of Technology}\\
{\footnotesize\it Ibaraki 305-0005, Japan}}

\date{}

\maketitle

\begin{abstract}

We show that zero-energy flows appear in many particle systems as same as 
in single particle cases in 2-dimensions. 
Vortex patterns constructed from the zero-energy flows can be investigated 
in terms of the eigenstates 
in conjugate spaces of Gel'fand triplets. 
Stable patterns are written by the superposition of zero-energy eigenstates. 
On the other hand 
vortex creations and annihilations are described 
by the insertions of unstable eigenstates with complex-energy eigenvalues 
into the stable patterns. 
Some concrete examples are presented in 
the 2-dimensional parabolic potential barrier case. 
We point out three interesting properties of the zero-energy flows; 
(i) the absolute economy as for the energy consumption, 
(ii) the infinite variety of the vortex patterns, and 
(iii) the absolute stability of the vortex patterns . 


\end{abstract}

\thispagestyle{empty}

\setcounter{page}{0}

\pagebreak
\hfil\break
{\bf 1. Introduction}

Vortices play interesting roles in various aspects of 
present-day physics such as vortex matters (vortex lattices) 
in condensed matters~\cite{blat,crab}, quantum Hall effects
 [3-5], 
various vortex patterns of 
non-neutral plasma [6-9] 
and Bose-Einstein gases [10-14]. 
Some fundamental properties and applications of vortices 
in quantum mechanics 
were examined by many authors [15-23]. 

Recently we have proposed a way to investigate vortex patterns 
in terms of zero-energy solutions of Schr\"{o}dinger 
equations in 2-dimensions, which are infinitely degenerate and 
eigenfunctions 
in conjugate spaces of Gel'fand triplets (CSGT)~\cite{k1,ks8}. 
It should be noted that the eigenfunctions in CSGT represent 
scattering states, and thus they are generally not normalizable [26]. 
Therefore, the probability density ($|\psi|^2$) and 
the probability current 
(${\vj}={\rm Re}[\psi^*(-i\hbar \nabla)\psi]/m$) for the 
eigenfunction ($\psi$), which are defined  
in usual Hilbert spaces, cannot be introduced to the eigenfunctions 
in CSGT. 
Instead of the probability current, however, the velocity 
which is defined by ${\vv}={\vj}/|\psi|^2$ can have a 
well-defined meaning, because the ambiguity due to the 
normalization of the eigenfunctions disappears in the definition 
of the velocity. 
Actually we have shown that many interesting objects used in hydrodynamics 
such as the complex velocity potential can be introduced 
in the 2-dimensions of CSGT [27]. 
We can expect that the hydrodynamical approach is a quite hopeful 
framework in the investigation of phenomena described in CSGT. 
One should pay attention to 
two important facts obtained in the early works [24,25]. 
One is the fact that the zero-energy solutions are common 
over the two-dimensional central potentials such that  
$V_a(\rho)=-a^2g_a\rho^{2(a-1)}$ 
with $\rho=\sqrt{x^2+y^2}$ except $a=0$ 
and then similar vortex patters described by the zero-energy solutions 
appear in all such potentials. 
Actually zero-energy solutions for a definite number of $a$ 
can be 
transformed to solutions for arbitrary number of $a$ by conformal 
transformations~\cite{k1,ks8}. 
The other is that the zero-energy solutions are infinitely degenerate, 
and thus all energy eigenvalues in CSGT with the potentials 
$V_a(\rho)$ are infinitely degenerate because the 
addition of the infinitely degenerate zero-energy solutions 
to arbitrary eigenfunctions does not 
change the energy eigenvalues at all. 

We, however, have to know that 
these results are obtained in equations for the single particle. 
We can, of course, apply the results in scattering processes where 
injected particles can be treated as individual particles. 
In the above-mentioned processes where vortices are observed [1-14], however, 
correlations among constituent particles cannot be ignored. 
We have to study whether such zero-energy flows appear in many particle systems. 
Furthermore we also have to study time evolutions of vortex patterns that 
have already been observed in experiments [6-14]. 
As for the time evolutions 
we have to take account of non-zero energy solutions in CSGT. 
It is known that the non-zero energy solutions 
generally have complex energy eigenvalues like ${\cal E}=E \mp i \Gamma$ 
with $E$, $\Gamma \in \real$,   
and then they have the time developments 
described by the factors $e^{-(i E/\hbar \pm \Gamma/\hbar)t}$~\cite{bohm}. 
Considering that stationary flows and time-dependent flows 
are, respectively, represented by the zero-energy solutions and the non-zero 
energy solutions in CSGT, 
we can expect that in CSGT general time-dependent flows 
are described by 
linear combinations of the stationary flows 
written by the zero-energy 
solutions and non-stationary flows (time-dependent flows) 
described by non-zero (complex) energy solutions. 
In hydrodynamics it is well-known that the vortex patterns are important 
objects to identify the situations of the flows.
In the present model 
we can image two different types of the vortex patterns. 
One is the stationary vortex patterns,  
and the other is the vortex patterns varying in the time evolutions. 
As presented in our early works ~\cite{k1,ks8}, the stationary vortex patterns can be 
described by the superposition of the zero-energy solutions, 
while the time-dependent ones will be done by putting the non-zero energy solutions 
into the superposition.

In this paper we shall study two problems; 
one is the zero-energy flows in many particle systems in section 2, 
and the other is time-developments of vortex patterns in section 3.
In order to obtain concrete examples of time-dependent vortex patterns, 
we shall use the eigenfunctions of 
the 2-dimensional parabolic potential barrier (2D-PPB) [27] in section 3,  
because the eigenfunctions with non-zero energy (complex energy) solutions 
in CSGT are known only in the case of the PPB. 
It is, however, noticed that the stable vortex patterns obtained in the 2D-PPB 
can easily be transformed those of the potentials $V_a(\rho)$ by the 
conformal transformations [24,25]. 
From these concrete analyses we shall point out three interesting properties of 
zero-energy flows in section 4. 
In section 5 some remarks will be done. 
Throughout these investigations 
we shall see that this approach can be one interesting possibility to analyze 
various time-dependent vortex patterns in a rigorous framework of quantum mechanics 
in CSGT.

\hfil\break
{\bf 2. Zero-energy flows and vortices}

\hfil\break
{\bf 2.1 Cases of single particle motions}

Let us briefly see the arguments for the zero-energy flows in the single 
particle motions. 
(For details, see refs. 24 and 25.)
\hfil\break
{\bf (1) Zero-energy flows in single particle motions}

In 2-dimensions 
the eigenvalue problems with the energy eigenvalue ${\cal E}$ 
are explicitly written by 
\begin{equation}
 [-{\hbar^2 \over 2m}\vtri +V_a(\rho)]\ \psi(x,y) 
  = {\cal E}\ \psi(x,y),
  \label{1}
\end{equation}
where 
$
\vtri=\partial^2/ \partial x^2+\partial^2 / \partial y^2,
$  
 the central potentials are generally given by  
$
V_a(\rho)=-a^2 g_a\rho^{2(a-1)},
$ 
with $\rho=\sqrt{x^2+y^2}$, $a \in \real$ ($a\not=0$), and 
$m$ and $g_a$ are, respectively, the mass of the particle and the coupling 
constant. 
Note here that the eigenvalues ${\cal E}$ should generally 
be complex numbers in CSGT. 

Let us consider the conformal mappings 
$ 
\zeta_a=z^a,\ \ \ \ \ \  {\rm with}\ z=x+iy.
$ 
We use the notations $u_a$ and $v_a$ defined by 
$
\zeta_a=u_a+iv_a
$ 
that are written as  
$ 
u_a=\rho^a\cos a\varphi,\ \ v_a=\rho^a\sin a\varphi,
$ 
where $\varphi=\arctan (y/x)$. 
In the $(u_a,v_a)$ plane the equations \eqref{1} are written down as 
\begin{equation}
a^2\rho_a^{2(a-1) / a} [-{\hbar^2 \over 2m}\vtri_a-g_a]\ \psi(u_a,v_a)=
     {\cal E}\ \psi(u_a,v_a), 
 \label{2}
\end{equation}
where 
$ 
\vtri_a=\partial^2/ \partial u_a^2+\partial^2/ \partial v_a^2.
$ 
We see that the equations become same for all values of $a$ 
(except $a=0$) for 
the energy eigenvalue ${\cal E}=0$.   
In fact, for ${\cal E}=0$ the equations have the same form as that for 
the free particle  with the constant potentials $g_a$ as 
\begin{equation}
[-{\hbar^2 \over 2m}\vtri_a-g_a]\ \ \psi(u_a,v_a)=0. 
 \label{3}
\end{equation} 
It should be noticed that in the case of $a=1$ where the original potential 
is a constant $g_1$ the energy does not need to be zero but can take arbitrary 
real numbers, because the right-hand side of \eqref{1} has no $\rho$ dependence.  
In the $a=1$ case, therefore, we should take $g_1+{\cal E}$ instead of 
$g_a$. 
This means that all plane wave solutions have the same infinite degeneracy 
discussed below. 

It is trivial that the equations for all $a$ have the 
particular solutions 
$ 
\psi_0^\pm(u_a)=N_a e^{\pm ik_au_a}
$ 
and 
$\psi_0^\pm(v_a)=N_a e^{\pm ik_av_a} 
$ 
with $k_a=\sqrt{2mg_a}/\hbar$ for $g_a>0. $ 
We notice here the degeneracy of the solutions that have already been known 
in the 2D-PPB [27].  
By putting the wave function $f^\pm (u_a;v_a)\psi_0^\pm(u_a)$ 
into~\eqref{3} 
where $f^\pm (u_a;v_a)$ is a polynomial function of $u_a$ and $v_a$, 
we obtain the equation 
\begin{equation}
[\vtri_a \pm2ik_a{\partial \over \partial u_a}]f^\pm (u_a;v_a)=0.
\label{4}
\end{equation} 
A few examples of the functions $f$ 
are given by 
\begin{align}
f_0^\pm (u_a;v_a)&=1, \nonumber \\
f_1^\pm (u_a;v_a)&=4k_a v_a, \nonumber \\
f_2^\pm (u_a;v_a)&=4(4k_a^2 v_a^2+1\pm 4 i k_a u_a).
\label{5}
\end{align} 
We can obtain the general forms of the polynomials in the 2D PPB, 
which are generally written by the multiple of 
the polynomials of degree $n$, $H_n^\pm(\sqrt{2k_2} x)$, 
such that 
$ 
f_{n}^{\pm}(u_2;v_2)=H_{n}^\pm(\sqrt{2k_2} x)
\cdot H_{n}^\mp(\sqrt{2k_2} y),
$ 
where $x$ and $y$ in the right-hand side should be 
considered as the functions of $u_2$ and $v_2$~\cite{sk4}. 
Since the form of the equations~\eqref{4}  
is common for all $a$, 
the solutions can be written by the same polynomial functions 
that are obtained in the PPB. 
The states expressed by these wave functions belong to the conjugate 
spaces of Gel'fand triplets of which nuclear space is given by 
Schwarz space. 
Actually we easily see that the wave functions cannot be normalized 
in terms of Dirac's delta functions except the lowest polynomial 
solutions. 

Here it should be stressed that 
the existence of the infinitely degenerate zero-energy solutions brings 
the infinite degeneracy to all the eigenstates. 
This fact means that the energy and the other quantum numbers 
like angular momentums, which are related to 
the determination of the energy eigenvalues, 
are not enough to discriminate the eigenstates. 
What are good quantum numbers to characterize the infinite degeneracy? 
An interesting candidate to characterize the states is vortex patterns 
that tell us topological properties of the staes.  
Note that the voertex patterns have been observed in experiments [8-14]. 
Time developments of the patterns can also be good observables 
in those processes. 
\vskip10pt
\hfil\break
{\bf (2) Velocity and vortices in quantum mechanics}

Let us here briefly note 
how vortices are interpreted in quantum mechanics. 
The probability density $\rho(t,x,y)$ and 
the probability current $\vj(t,x,y)$ of a wavefunction $\psi(t,x,y)$ 
in non-relativistic quantum mechanics 
are, respectively, defined 
by 
$ 
  \rho(t,x,y)\equiv\left| \psi(t,x,y)\right|^2
  $ and 
  $
  \vj(t,x,y)\equiv{\rm Re}\left[\psi(t,x,y)^*
  \left(-i\hslash\nabla\right)\psi(t,x,y)\right]/m.
$ 
They satisfy the equation of continuity 
 $ \partial\rho/\partial t+\nabla\cdot\vj=0.$ 
Following the analogue of the hydrodynamical 
approach, 
the fluid 
can be represented by 
the density $\rho$ and the fluid velocity $\vv$. 
They satisfy Euler's equation of continuity 
$ 
  \partial\rho/\partial t+\nabla\cdot(\rho\vv)=0. 
$ 
Comparing this equation with the continuity equation, 
the following 
definition for the quantum velocity of the state $\psi(t,x,y)$ 
is led in the hydrodynamical approach; 
\begin{equation}
  \vv\equiv\frac{\vj(t,x,y)}{\left| \psi(t,x,y)\right|^2}. 
\end{equation}
Now it is obvious that vortices appear at the zero points of 
the density, that is, the nodal points of the wavefunction.  
At the vortices, of course, 
the current $\vj$ must not vanish. 
When we write the wavefunction $\psi(t,x,y)=\sqrt{\rho (x,y)} e^{iS(x,y)/\hbar }$, 
the velocity is given by ${\vv}=\nabla S/m$. 
We should here remember that the solutions 
degenerate infinitely. 
This fact indicates that we can construct wavefunctions having 
the nodal points at arbitrary positions in terms of 
linear combinations of the infinitely degenerate 
solutions~\cite{k1,ks8,sk4}. 

The strength of vortex is characterized 
by the circulation $\Gammav$ 
that is represented by the integral round a closed contour $C$
encircling the vortex such that 
\begin{equation}
\Gammav=\oint_C \vv \cdot d\vs
\end{equation} 
and it is quantized as 
\begin{equation}
\Gammav=2\pi l\hbar/m, 
\end{equation} 
where the circulation number $l$ is 
an integer~\cite{joh2,joh5,wu-sp,bb2}.

\hfil\break 
{\bf 2.2 Cases of many particle motions}

Let us consider a simple system composed of $N$ number of the same 
particles with the mass $m$. 
\hfil\break
{\bf (1) Zero-energy flows in many particle cases}

The interactions between two constituent particles are supposed to be 
written by the same 
potential $V(\rho_{ij})$, where $\rho_{ij}=|\vec{\rho_i}-\vec{\rho_j}| $ 
stands for the relative distance between two particles. 
The Schr$\ddot o$dinger equation for the $N$ particle system is written as 
\begin{equation} 
[-{\hbar^2 \over 2m}\sum_{i=1}^N\vtri_i 
+\sum_{i>j}^N \sum_{j=1}^{N-1}V(\rho_{ij})]\ \Psi(t,\vec{\rho}_1,\cdots,\vec{\rho}_N) 
  = {\cal E}\ \Psi(t,\vec{\rho}_1,\cdots,\vec{\rho}_N).
\end{equation}
Introducing a centre of mass coordinate $\vec{\rho}_C$ and $N-1$ relative coordinates 
$\vec{\rho}_{r_i}$ with $i=1,\cdots,N-1$, the equation is rewritten as 
\begin{equation} 
[-{\hbar^2 \over 2M}\vtri_C-H_r(\rho_{r_1},\cdots,\rho_{r_{N-1}})]
\Phi_C(t,\vec{\rho_C})\phi_r (t,\vec{\rho}_{r_1},\cdots,\vec{\rho}_{r_{N-1}}) 
  = {\cal E}\ \Phi_C(t,\vec{\rho}_C)\phi_r (t,\vec{\rho}_{r_1},\cdots,\vec{\rho}_{r_{N-1}}),
\end{equation}
where $M=Nm$ and $H_r(\rho_{r_1},\cdots,\rho_{r_{N-1}})$ 
stands for the Hamiltonian for the relative coordinates. 
It should be noted that the centre of mass coordinate can always be 
separable from the relative ones, and then the energy eigenvalue ${\cal E}$ 
are written by the sum of the eigenvalue of centre of mass system $E_C$ 
and that of the relative ones $E_r$ 
as ${\cal E}=E_C +E_r$. 
We can consider that the zero-energy solutions can appear in the relative motions, 
e.g., when the relative interactions are written by the same PPB, $H_r$ are separable 
for all the relative coordinates, for which interatcions are written by the PPB. 
Here we shall, however, discuss  the case where the flows discussed in the section 2.1  
appear in the centre of mass motions, that is, the total flows. 
Now let us consider the zero-energy solutions of the total system 
represented by ${\cal E}=0$. 
Considering the equation ${\cal E}=E_C +E_r$, 
the flows discussed in the section 2.1 appear for $E_r<0$, because $E_C>0$ 
is required. 
This means that the relative interactions have to be totally described 
by a kind of attractive potential. 
In this case the centre of mass motions $\Phi_C(t,\vec{\rho_C})$ 
are described by the plane waves with the 
infinite degeneracy, which are obtained from the solutions in the section 2.1 
by putting $a=1$. 
If the Hamiltonian $H_r$ has levels with different negative-energy eigenvalues, 
we have different zero-energy flows characterized by different wave numbers 
$k=\sqrt{2M|E_r|}/\hbar $ corresponding to the negative-energy eigenvalues $E_r<0$. 

Provided that some external forces like electro-magnetic forces are put in the systems, 
the central motions possibly have potentials ($V_C(\rho_C)$). 
In this case the zero-energy solutions appear when $V_C(\rho_C)$ is written by one of 
the central potentials $V_a(\rho_C)$. 

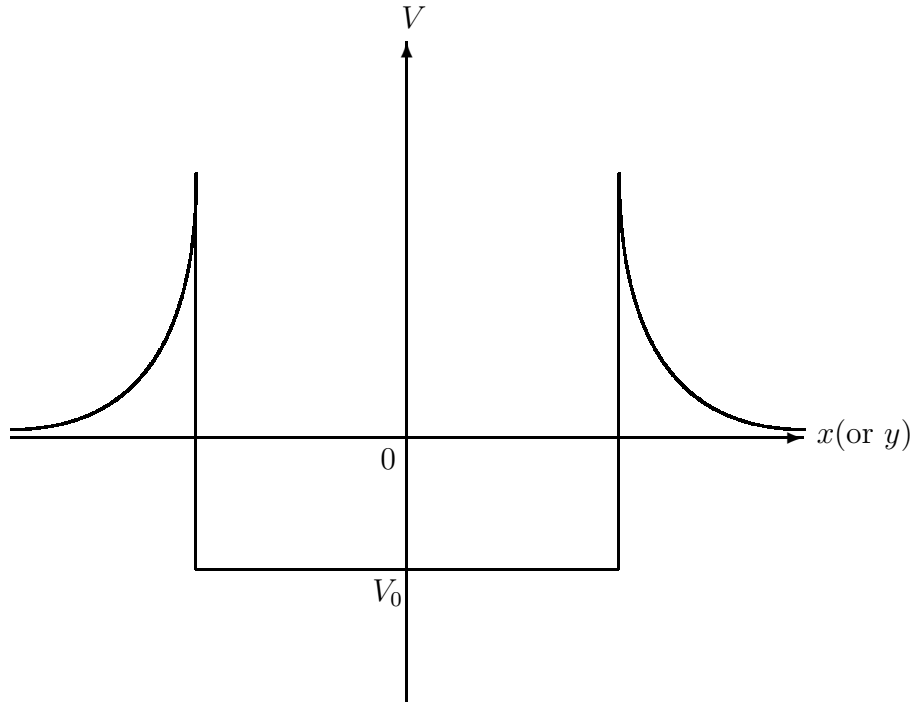
\begin{figure}[pb]
   \begin{center}
    \begin{picture}(300,300)
     \thicklines
     \put(0,100){\vector(1,0){300}}
     \put(150,0){\vector(0,1){250}}
     \put(140,88){$0$}
     \put(305,98){$x({\rm or}\ y)$}
     \put(148,255){$V$}
     
     \qbezier(230,200)(230,103)(300,103)
     \qbezier(70,200)(70,103)(0,103)
     
     \put(70,50){\line(0,1){150}}
     \put(230,50){\line(0,1){150}}
     \put(70,50){\line(1,0){160}}
     \put(137,38){$V_0$}
     
    \end{picture}
   \end{center}
   \caption[]{An example of the potentials with zero-energy solutions.}
   \label{fig:1.1}
  \end{figure}

Here we note that 
in cases where 
the potentials of all the constituents  
have a negative constant potential 
$V_0<0$ in a common area such as 
in the region surrounded by repulsive potentials 
like short distance attractive potentials 
as shown in fig.~\ref{fig:1.1}, 
the zero-energy solutions can appear in the area. 
In these cases we have to solve the zero-energy problems under some special 
boundary conditions. 
We will be able to find out the solutions fulfilling the boundary conditions 
in terms of the infinitely degenerate zero-energy solutions. 
\hfil\break
{\bf (2) Velocity of the centre of mass system}  

It is obvious that we can introduce the velocity for the centre of mass system 
as same as that for the single particle given in section 2.1 such that 
\begin{equation} 
\vv_C=\vj_C(t,\vec{\rho_C})/|\Phi_C(t,\vec{\rho_C})|^2, 
\end{equation} 
where the current for the centre of mass system is defined by 
 $
  \vj_C(t,\vec{\rho}_C)=
 \break
 {\rm Re}\left[\Phi_C^*
  \left(-i\hslash\nabla_{\rho_C}\right)
  \Phi_C\right]/m.
$ 
Since the derivative with respect to the centre of mass coordinate is just written 
by the sum of the derivatives of all the constituents, 
the velocity $\vv_C$ is understood as the mean velocity of the constituents. 
We may consider that $\vv_C$ represents the velocity that is used 
in hydrodynamics. 
Now it is trivial that in the case of the centre of mass motions 
we can follow the discussion on the vortices for the single particle motions 
presented in section 2.1.

We note that the velocity for a relative coordinate $\rho_i$ can also be 
defined as same as that for the centre of mass coordinate such that 
\begin{equation}
\vv_i=\vj_{\rho_i}/|\phi_r|^2,
\end{equation} 
where 
$
  \vj_{\rho_i}={\rm Re}\left[\phi_r^*
  \left(-i\hslash\nabla_{\rho_i}\right)\phi_r \right]/m.
$ 
We easily see that in independent particle models the above velocity 
coincides with that of the single particle given in section 2.1. 
If the central potentials of the type $V_a(\rho)$ appear 
for some relative coordinates, the zero-energy flows appear, and 
then the zero-energy vortices are produced in the relative motions.

\hfil\break
{\bf 3. Vortex patterns in 2D-PPB}

In order to see some concrete examples of stable and 
time-dependent vortex patterns we shall make some vortex patterns 
in terms of the eigenfunctions of the 2D-PPB [27]. 
As already noted, the stable patterns can be transformed to those of 
arbitrary potentials of the type $V_a(\rho)$ by the conformal 
transformations [24,25]. 

\hfil\break
{\bf 3.1 Eigenfunctions of 2D-PPB}

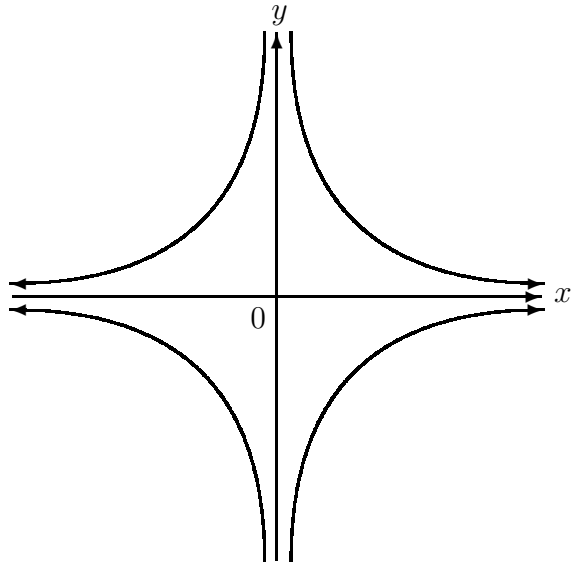
\begin{figure}[pb]
   \begin{center}
    \begin{picture}(200,200)
     \thicklines
     \put(0,100){\vector(1,0){200}}
     \put(100,0){\vector(0,1){200}}
     \put(90,88){$0$}
     \put(205,98){$x$}
     \put(98,205){$y$}
     
     \qbezier(105,200)(105,105)(200,105)
     \qbezier(95,200)(95,105)(0,105)
     \qbezier(95,0)(95,95)(0,95)
     \qbezier(105,0)(105,95)(200,95)
     
     \put(200,105){\vector(1,0){1}}
     \put(0,105){\vector(-1,0){1}}
     \put(0,95){\vector(-1,0){1}}
     \put(200,95){\vector(1,0){1}}
    \end{picture}
   \end{center}
   \caption[]{Corner flows moving from the $y$-direction 
   to the $x$-direction.}
   \label{fig:1.2}
  \end{figure}

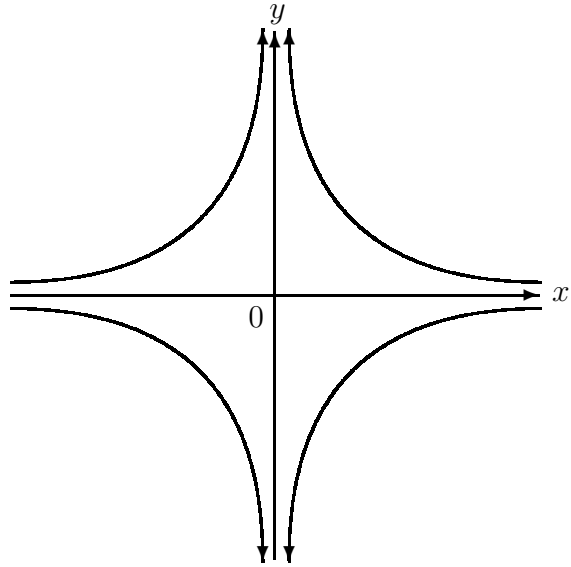
\begin{figure}
   \begin{center}
    \begin{picture}(200,200)
     \thicklines
     \put(0,100){\vector(1,0){200}}
     \put(100,0){\vector(0,1){200}}
     \put(90,88){$0$}
     \put(205,98){$x$}
     \put(98,205){$y$}
     
     \qbezier(105,200)(105,105)(200,105)
     \qbezier(95,200)(95,105)(0,105)
     \qbezier(95,0)(95,95)(0,95)
     \qbezier(105,0)(105,95)(200,95)
     
     \put(105.5,200){\vector(0,1){1}}
     \put(95.5,200){\vector(0,1){1}}
     \put(95.5,0){\vector(0,-1){1}}
     \put(105.5,0){\vector(0,-1){1}}
    \end{picture}
   \end{center}
   \caption[]{Corner flows moving from the $x$-direction 
   to the $y$-direction.}
   \label{fig:1.3}
   \end{figure}

Let us start from the short review on the eigenfunctions of the 2D-PPB 
$V=-m\gamma^2(x^2+y^2)/2$ that 
will be used in the following analyses. 
The infinitely degenerate eigenfunctions with the zero-energy for the potentials 
$V_a(\rho)$ 
have explicitly been given in (5), whereas the eigenfunctions 
with non-zero energies are known only in the case of PPB~
[27-33]. 
It is trivial that the eigenfunctions in the 2D-PPB are represented by the multiples of 
those of the 1D-PPB.
The eigenfunctions of the 1D-PPB for $x$, which have pure imaginary energy eigenvalues 
$ \mp i( n_x+{1 \over 2})\hbar \gamma$, are given by 
\begin{equation}
u^{\pm}_{n_x}(x)=e^{\pm i\beta^2x^2/2}H^{\pm}_{n_x}(\beta x)\ \ \ \ 
(\beta \equiv \sqrt{m\gamma/\hbar}),
\end{equation} 
where $ H^{\pm}_{n_x}(\beta x)$ are the polynomials of degree $n_x$ written 
in terms of Hermite polynomials 
$H_{n}(\xi)$ with $\xi =\beta x$ as~\cite{sk,s2} 
\begin{equation}
H_{n}^\pm(\xi)=e^{\pm i n\pi/4} H_{n}(e^{\mp i\pi/4}\xi). 
\end{equation}
We have four different types of the eigenfunctions in the 2D-PPB~\cite{sk4}. 
Two of them 
$$U^{\pm\pm}_{n_xn_y}(x,y)\equiv u^{\pm}_{n_x}(x)u^{\pm}_{n_y}(y)$$  
with the energy eigenvalues ${\cal E}^{\pm\pm}_{n_xn_y}=\mp i(n_x+n_y+1)$, 
respectively, represent flows diverging from the origin 
and flows converging towards the origin. 
Some examples for the low degrees are obtained as follows; 
\begin{align}
U^{\pm\pm}_{00}(x,y)&=e^{\pm i\beta^2(x^2+y^2)/2} ,  \nonumber \\
U^{\pm\pm}_{10}(x,y)&=2\beta x e^{\pm i\beta^2(x^2+y^2)/2} , \nonumber \\ 
U^{\pm\pm}_{20}(x,y)&=(4\beta^2x^2 \mp 2i) e^{\pm i\beta^2(x^2+y^2)/2}, \nonumber \\
U^{\pm\pm}_{11}(x,y)&=4\beta^2xy e^{\pm i\beta^2(x^2+y^2)/2} .
\end{align}
Note that $U^{\pm\pm}_{01}(x,y)$ and $U^{\pm\pm}_{02}(x,y)$ are ,respectively, 
obtained by exchanging $x$ and $y$ in $U^{\pm\pm}_{10}(x,y)$ and 
$U^{\pm\pm}_{20}(x,y)$. 
It is transparent that the eigenfunctions of angular momentums are constructed 
in terms of the linear combinations of these diverging and converging 
flows~\cite{sk4}. 
The other two  
$$U^{\pm\mp}_{n_xn_y}(x,y)\equiv u^{\pm}_{n_x}(x)u^{\mp}_{n_y}(y)\ \ {\rm with}\ \  
{\cal E}^{\pm\mp}_{n_xn_y}(x,y)=\mp i(n_x-n_y)$$ 
are corner flows round the center. 
(See figs.~\ref{fig:1.2} and~\ref{fig:1.3}.) 
The zero-energy solutions that are common in the potentials $V_a(\rho)$ 
appear when $n_x=n_y$ is satisfied. 
A few examples of the zero-energy eigenfunctions 
are explicitly obtained as follows; 
\begin{align}
U^{\pm\mp}_{00}(x,y)&=e^{\pm i\beta^2(x^2-y^2)/2} ,  \nonumber \\
U^{\pm\mp}_{11}(x,y)&=4\beta^2 xy e^{\pm i\beta^2(x^2-y^2)/2} , \nonumber \\
U^{\pm\mp}_{22}(x,y)&=4[4\beta^4x^2 y^2+1\pm 2i\beta^2(x^2-y^2)]
              e^{\pm i\beta^2(x^2-y^2)/2}.
\end{align}
It is obvious that the eigenfunctions of (15) and (16) are not normalizable. 
The proof that they are the eigenfunctions of CSGT are presented 
in refs. [32,33].

\hfil\break
{\bf 3.2 Stable vortex-patterns}

Let us investigate vortex patterns in terms of the 2D-PPB eigenfunctions. 
Since the zero-energy solution given in the section 3.1 have no nodal point 
with non-vanishing currents, 
they have no vortex. 
However, it has been shown that 
some vortex patterns having infinite numbers of vortices, 
like vortex lines 
and vortex lattices, can be made in terms of simple linear combinations of 
those low lying stationary states in the early works~\cite{k1,ks8}. 
We shall here study linear combinations having a few or some vortices 
observed in experiments [6-14]. 
Let us start from compositions 
of stable vortex patterns. 
For the convenience in the following discussions 
we take the flow without any vortices described by 
\begin{align} 
\Phi_B&={1 \over 4} U_{22}^{+-}(x,y) -U_{00}^{+-}(x,y)    \nonumber \\
      &=(4\beta^4x^2y^2+2i\beta^2(x^2-y^2))e^{i\beta^2(x^2-y^2)/2}, 
\end{align} 
which will be called the basic flow hereafter. 
Note that the nodal point of the basic flow 
at the origin does not produce vortices, 
because the current vanishes there.

Here we shall show three 
stable patterns, which will be 
used in the discussions on vortex creations and  annihilations.
\hfil\break 
{\bf (S-1) Three-vortices pattern}

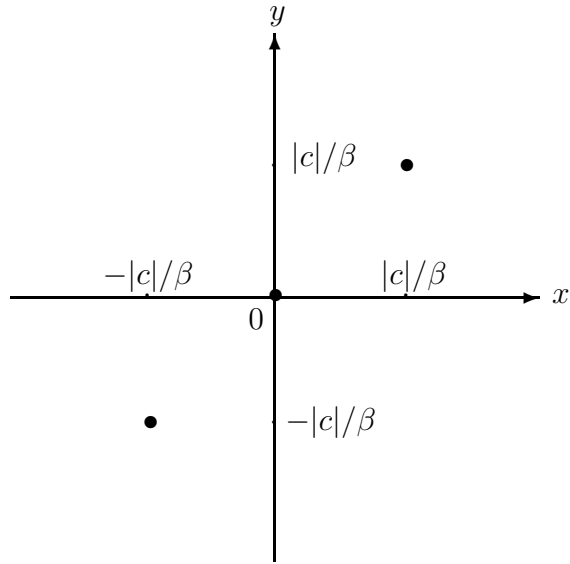
\begin{figure}
    \begin{center}
    \begin{picture}(200,200)
     \thicklines
     \put(0,100){\vector(1,0){200}}
     \put(100,0){\vector(0,1){200}}
     \put(90,88){$0$}
     \put(205,98){$x$}
     \put(98,205){$y$}
     
     \put(148,98){$\cdot $}
     \put(50,98){$\cdot $}
     \put(140,105){$|c|/\beta $}
     \put(35,105){$-|c|/\beta $}
     \put(98,147){$\cdot $}
     \put(98,50){$\cdot $}
     \put(106,150){$|c|/\beta $}
     \put(104,50){$-|c|/\beta $}

     \put(147,147){$\bullet$}
     \put(97.4,97.8){$\bullet$ }
     \put(50,50){$\bullet $}
    \end{picture}
   \end{center}
   \caption[]{Stationary pattern with three vortices.
        $\bullet$ denotes a vortex.}
   \label{fig:1.4}
    \end{figure}

The linear combination of the basic flow 
and $U^{+-}_{11}$ such that 
\begin{align}
\Phi^{+-}_{012}(x,y)&=\Phi_B-c^2 U^{+-}_{11}(x,y) \nonumber \\
             &=[4\beta^2xy(\beta^2xy-c^2)+2i\beta^2(x^2-y^2)]
             e^{i\beta^2(x^2-y^2)/2},
\end{align}
with $c \in \real$ 
has three vortices at the origin and the two points $( \pm c/\beta, \pm c/\beta)$ 
as shown in fig.~\ref{fig:1.4}. 
If we take $-c^2$ instead of $c^2$, three vortices appear at the origin and the points 
$( \pm c/\beta, \mp c/\beta)$.
\hfil\break 
{\bf (S-2) One-vortex pattern}

Let us consider 
the linear combinations of two flows that are represented by 
the above eigenfunction $\Phi^{+-}_{012}(x,y)$ and 
coming from two different directions such that 
\begin{align}
\Phi^{+-}_{012}(x,y)+\Phi^{+-}_{012}(\xi,\eta),
\end{align}
where $\xi={\rm cos}\alpha \cdot x+{\rm sin}\alpha \cdot y$ and 
$\eta=-{\rm sin}\alpha \cdot x+{\rm cos}\alpha \cdot y $ 
with $0<\alpha <2\pi $. 
It is apparent that the two functions have only one common zero-point at 
the origin for $0<\alpha <2\pi $. 
We see that the vortex at the origin has the circulation number 
$l=-2$. 
\hfil\break
{\bf (S-3) Four-vortices pattern }

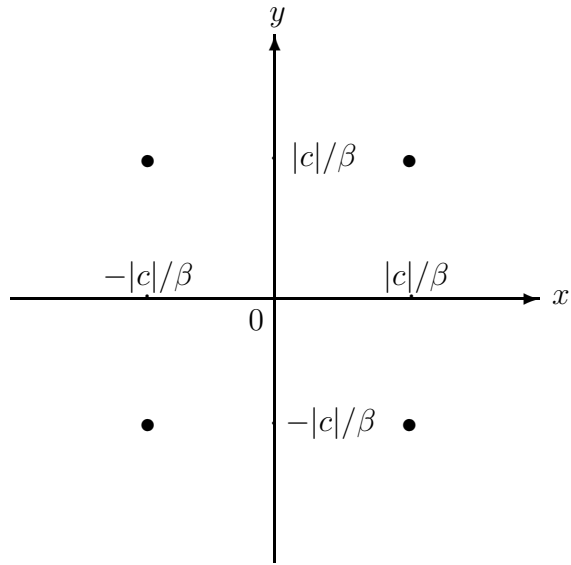
\begin{figure}
    \begin{center}
    \begin{picture}(200,200)
     \thicklines
     \put(0,100){\vector(1,0){200}}
     \put(100,0){\vector(0,1){200}}
     \put(90,88){$0$}
     \put(205,98){$x$}
     \put(98,205){$y$}
     
     \put(150,98){$\cdot $}
     \put(50,98){$\cdot $}
     \put(141,105){$|c|/\beta $}
     \put(35,105){$-|c|/\beta $}
     \put(98,150){$\cdot $}
     \put(98,50){$\cdot $}
     \put(106,150){$|c|/\beta $}
     \put(104,50){$-|c|/\beta $}

     \put(148,149){$\bullet $}
     \put(148,49){$\bullet $}
     \put(49,149){$\bullet $}
     \put(49,49){$\bullet $}
    \end{picture}
   \end{center}
   \caption[]{Stationary pattern with four vortices.}
   \label{fig:1.5}
    \end{figure}

The linear combinations given by 
\begin{align}
\Phi^{+-}_{02}(x,y)&={1 \over 4}\Phi_B-c^4U^{+-}_{00}(x,y) \nonumber \\
             &=[(\beta^2xy-c^2)(\beta^2xy+c^2)+i\beta^2(x^2-y^2)/2]
             e^{i\beta^2(x^2-y^2)/2},
\end{align} 
has four vortices 
at the points $( \pm c/\beta, \pm c/\beta)$ and 
$( \pm c/\beta, \mp 1c/\beta)$ as shown in fig.~\ref{fig:1.5}.


\hfil\break
{\bf 3.3 Moving vortex-patterns} 

Let us go to the study of time-dependent vortex patterns, 
where vortices can move, and sometimes be created and annihilated. 
Such patterns are obtained by linear combinations of 
stable flows and time-dependent ones. 
In the following considerations the time dependent flows are put in 
the stable flows at $t=0$. 
\hfil\break
{\bf (M-1) A pattern accompanied by a pair creation and annihilation}

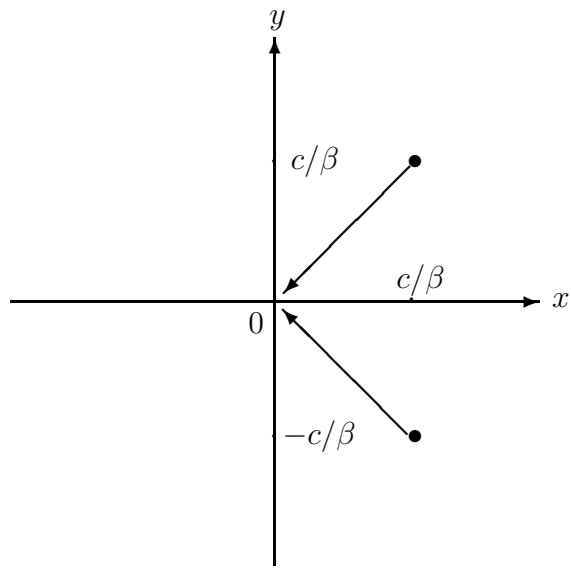
\begin{figure}
    \begin{center}
    \begin{picture}(200,200)
     \thicklines
     \put(0,100){\vector(1,0){200}}
     \put(100,0){\vector(0,1){200}}
     \put(90,88){$0$}
     \put(205,98){$x$}
     \put(98,205){$y$}
     
      \put(150,150){$\bullet$}
     \put(150,46){$\bullet $}
     
     \put(150,98){$\cdot $}
     \put(146,105){$c/\beta $}
     \put(98,150){$\cdot $}
     \put(98,46){$\cdot $}
     \put(106,150){$c/\beta $}
     \put(103,46){$-c/\beta $}
     
     \put(151,151){\vector(-1,-1){47.65}}
     \put(150,50){\vector(-1,1){47}}
    \end{picture}
   \end{center}
   \caption[]{A pair annihilation pattern for $c>0$. 
   The arrows show the moving directions of the vortices.}
   \label{fig:1.6}
  \end{figure}

Let us consider the following linear combination; 
\begin{align}
\Phi^{+-}_{02,1}(x,y,t)&={1 \over 2}\Phi_B-\theta(t)c^3U^{+-}_{10}(x,y)
          e^{-\gamma t} \nonumber \\
             &=[2\beta x(\beta^3xy^2-\theta(t)c^3e^{-\gamma t})+i\beta^2(x^2-y^2)]
          e^{i\beta^2(x^2-y^2)/2}, 
\end{align} 
where the theta function is taken as 
$\theta (t)=0$ for $t<0$ and $=1$ for $t\geq 0$. 
It has two nodal points at 
$(c e^{-\gamma t/3}/\beta,\pm c e^{-\gamma t/3}/\beta)$ for $t\geq 0$, 
where two vortices with opposite circulation numbers  exist. 
The nodal points go to the origin as the time $t$ goes to infinity as shown 
in fig.~\ref{fig:1.6}. 
Since the contribution of the unstable flow 
decreases 
as $t \rightarrow \infty $ because of the time factor $e^{-\gamma t}$,  
the wavefunction 
$\Phi^{+-}_{02,1}(x,y,t)$ 
goes to $\Phi_B/2$ as $t \rightarrow \infty$. 
Thus the flow has no nodal point in the limit. 
This means that the pair of vortices which are created at $t=0$ 
disappears at origin in the limit $t \rightarrow \infty $. 
We can say that this wavefunction describes the pair annihilation of two vortices. 
The time development of this process can be described as follows: 
\hfil\break 
(i) Before the time-dependent flow is put in the basic flow, i.e., $t<0$, 
there is no vortex. 
\hfil\break 
(ii) At $t=0$ when the time-dependent flow is put in the basic flow, 
a pair of vortices are suddenly created. 
\hfil\break 
(iii) The pair moves toward the origin, and then they annihilate at the origin, 
that is, the flow turns back to the basic flow $\Phi_B$ having no vortex. 
\hfil\break
{\bf (M-2) A pattern accompanied by creations and annihilations of two pairs}

The linear combination given by 
\begin{align}
\Phi^{+-}_{02,2}(x,y,t)&=\Phi_B-\theta(t)c^2[2iU^{+-}_{00}(x,y)
                -U^{+-}_{20}(x,y)e^{-2\gamma t}] \nonumber \\
  &=[4\beta^2x^2(\beta^2y^2-\theta(t)c^2e^{-2\gamma t})-
         2i(\theta(t)c^2(1-e^{-2\gamma t})-\beta^2(x^2-y^2))]
             e^{i\beta^2(x^2-y^2)/2}
\end{align} 
has four nodal points at $(\pm c/\beta, c e^{-\gamma t}/\beta)$ and 
$(\pm c/\beta, -c e^{-\gamma t}/\beta)$ for $t\geq 0$. 
In this case we easily see that 
the pair of vortices at $(\pm c/\beta, c e^{-\gamma t}/\beta)$ and that 
at $(\pm c/\beta, -c e^{-\gamma t}/\beta)$ annihilate as $t \rightarrow \infty $ 
as shown in fig.~\ref{fig:1.7}.  
The stable flow $\Phi_B-2ic^2U^{+-}_{00}(x,y)$ that appears in the limit 
has two nodal points at $(\pm c/\beta,0)$, but it has no vortex but 
two vortex dipoles, because the 
current also vanish at the points. 
The time development of this process is interpreted similarly as the case (M-1). 

In these two cases all vortices move on straight lines in the pair annihilation 
processes.
\hfil\break
{\bf (M-3) A pattern accompanied by creation and annihilation of four vortices}

The linear combinations of stationary flows and diverging or converging 
ones make different types of annihilation processes. 
For an example, let us consider the linear combination of $ \Phi_B$ and 
the lowest order diverging flow described by 
\begin{equation}
U_{00}^{++}(x,y,t)=e^{i\beta^2(x^2+y^2)/2}e^{-\gamma t}
\end{equation}  
having the energy eigenvalue $-i \gamma \hbar$. 
Let us consider the linear combination described by 
\begin{align}
\Phi^{++}_{02,0}(x,y,t)&={1 \over 4}\Phi_B
                -\theta(t)c^2 U^{++}_{00}(x,y,t) \nonumber \\
  &=[\beta^4x^2y^2-\theta(t)c^2 e^{-\gamma t}e^{i\beta^2y^2}+i\beta^2(x^2-y^2)/2]
             e^{i\beta^2(x^2-y^2)/2}.
\end{align} 
For $t\geq 0$ it has two nodal points at the points where the following relations 
are fulfilled; 
\begin{equation}
XY=c(t) {\rm cos}Y,\ \ \ {1 \over 2}(X-Y)-c(t){\rm sin}Y=0, 
\end{equation} 
where $X=\beta^2 x^2$, $Y=\beta^2 y^2$ and $c(t)=c^2 e^{-\gamma t}$. 
From these relations we have an equation for $Y$  
\begin{equation}
Y^2-c(t) {\rm cos}Y +2c(t)Y{\rm sin}Y=0. 
\end{equation} 
The solutions are obtained from the cross points 
of two functions 
$f(Y)=Y^2$ and $g(Y)=c(t)({\rm cos}Y-2Y{\rm sin}Y)$. 
We easily see that a solution for $Y\geq 0$ exists in the region 
$0<Y<\pi /2$ for arbitrary positive numbers of $c(t)$. 
Four vortices appear at the four points expressed by the combinations of 
$x=\pm \sqrt{X}/\beta $ and $y=\pm \sqrt{Y}/\beta$, where 
$X$ is obtained by using the first relation of (25). 
Since $c(t)$ goes to $0$ as $t \rightarrow \infty $, we see that 
 $X$ and $Y$ 
simultaneously go to $0$ in the limit such that 
$$
X\simeq Y\rightarrow |c|e^{-\gamma t/2} \rightarrow 0, \ \ \ 
{\rm for}\ \  t \rightarrow \infty .  
$$ 
Since the flow turns back to the basic flow, the four vortices 
annihilate at the origin in the limit. 
From the second relation of (25), 
we have 
$$
X=Y+2c(t) {\rm sin}Y.
$$ 
This equation show us that the vortex points do not 
move along straight lines. 

Here we consider a somewhat complicated processe. 
\hfil\break
{\bf (M-4) A pattern accompanied by annihilation of four vortex-pairs} 

Here we take $\Phi^{+-}_{02}(x,y)$ of (20) as the stable flow, 
which has four vortices. 
For the simplicity $c=1/2$ is taken in the following discussions. 
Here the lowest order diverging flow 
$U_{00}^{++}(x,y,t)$ 
 are put into the stationary flow at $t=0$. 
The wavefunction are given by 
\begin{align}
\Phi_{02,0}^{++}(x,y;t)=&16\Phi^{+-}_{02}(x,y)+\theta(t)b^2U_{00}^{++}(x,y,t)  \nonumber  \\
                       =&[16\beta^4x^2y^2-1+\theta(t)b^2 e^{-\gamma t}e^{i\beta^2y^2}+
                      8i\beta^2(x^2-y^2)]
             e^{i\beta^2(x^2-y^2)/2} \nonumber \\
             =&[16\beta^4x^2y^2-1+\theta(t)b(t) e^{-\gamma t}{\rm cos}(\beta^2y^2)
               +i(8\beta^2(x^2-y^2)   \nonumber \\
             &+
             \theta(t)b(t) e^{-\gamma t}{\rm sin}(\beta^2y^2))]
             e^{i\beta^2(x^2-y^2)/2},
\end{align} 
where $b\in \real $ and $b(t) =b^2e^{-\gamma t}$. 
Using $X=\beta^2x^2$ and $Y=\beta^2y^2$, we have two relations for nodal points 
of the wavefunction for $t>0$ as follows;
\begin{equation} 
16X Y +b(t) {\rm cos}Y-1=0, \ \ \ \ 
8(X-Y)+b(t){\rm sin}Y=0.
\end{equation} 
From these relations we obtain an equation for the nodal points 
\begin{equation} 
1-b(t){\rm cos}Y -16Y^2+2b(t)Y{\rm sin}Y=0.
\end{equation} 
Examining the cross point of the two functions 
$F(Y)=16Y^2-1$ and $G(Y)=-b(t)({\rm cos}Y -2Y{\rm sin}Y)$, 
we obtain the following results:   
\hfil\break
(1) In the case of $b(t)<1$ 
the two functions always have a cross-point in the region satisfying 
$Y\geq 0$ (note that $Y=\beta^2y^2$). 
The wavefunction, therefore, has four nodal points. 
This means that the flow always has four vortices that move toward the stationary 
points fulfilling $|x|=|y|=(2 \beta)^{-1}$ as $t$ increases. 
\hfil\break
(2) In the case of $b(t)>1$, eq.(29) has an even number of solutions like 
$n=0,2,4,\cdots$. 
Since one solution brings four vortices on a circle with the centre at the origin, 
the vortex number is given by $4n$. 
Note that the number $n$ increases as $b(t)$ increases. 
This fact means that, since $b(t)$ decreases as $t$ increases, the vortex number 
decreases as $t$ increases, until $b(t)$ gets to 1. 
Considering that the change of $n$ is always 2, we see that 
the reduction of the vortex number caused by the change of $n$ is always 8. 
That is to say, we observe that four vortex-pairs simultaneously annihilate 
at four different points on a circle. 
As a simple example, let us consider the case of $n=2$ at $t=0$. 
We observe the following time development of the flow: 
   \hfil\break
(i) 
For $t<0$ the stationary flow has the four vortices as shown in fig.5. 
   \hfil\break
(ii) 
At $t=0$ the original four vortices are disappear and 
eight vortices are newly created. 
Then we observe 
the flow having eight vortices. (See fig.~\ref{fig:1.8}.) 
    \hfil\break
(iii) 
In the time-evolution the eight vortices disappear simultaneously. 
We observe the process as the annihilations of four vortex-pairs. 
Thus the flow having no vortex appears. 
    \hfil\break
(iv) 
At the critical time $t_c= {\rm ln} b^2 /\gamma$ 
when $b(t_c)=1$ is fulfilled  four vortices are created at the origin, and then 
they move toward the stationary points. 
The vortex state at $t=t_c$ can be understood as a vortex quadrupole~\cite{k1}. 

If, instead of the diverging flow  $U_{00}^{++}(x,y,t)$, 
the converging flow $U_{00}^{--}(x,y,t)$ is put in the stationary flow, 
we observe a flow continuously creating 8 vortices for any choices of $b$. 
Of course, the time dependent flow blows up the magnitude in the limit 
of $t\rightarrow \infty$, and thus the original stable flow can not be 
observed in the limit. 

\hfil\break
{\bf 4. Interesting properties of zero-energy flows}

Let us here consider interesting properties of the zero-energy flows. 
The first interesting property is due to the fact 
that the use of the zero-energy flows is very useful and economical 
from the viewpoint of energy consumption. 
For example it can be a very economical step  
for the transmission of information. 
Considering the huge variety arising from the infinite 
degeneracy, 
the transmission by the use of the zero-energy flows 
enable us to transmit an enormous 
amount of information without any energy loss. 
The flows are stable and then 
they can also be a very useful step for making mechanisms to preserve 
such information, e.g., for memories in living beings. 
The huge variety of vortex patterns can possibly discriminate 
the enormous amount of information.  
The addition of new memories 
and also the change of preserved memories can easily be carried out 
by pouring some 
zero-energy (stable) flows in the preserved ones. 
Furthermore, as shown in section 3.3, 
in all the time-dependent processes 
induced by pouring the unstable flows 
with complex energy eigenvalues the time-dependent flows always turn back 
to the stable flows in the long time scales, 
and then the initial flow patterns are recovered. 
That is to say, the initial patterns are kept in all such time-dependent processes. 
This stability of the flow patterns seems to be a very interesting property 
for the interpretation of the stability of memories not only in their preservations  
but in their applications as well. 
The applications, of course, mean thinking processes.     
These flows will possibly be workable in the steps for  
thinking in living beings. 
The use of the zero-energy and complex-energy solutions 
enables living beings to make up many functions in their bodies 
very economically on the basis of energy consumption.  
Anyway the zero-energy and complex-energy solutions are 
interesting objects to describe 
mechanisms working very economically as for the energy consumption. 
Especially, in the 2D-PPB case we can do it without any energy loss, 
because all the solutions of PPB have no real energy eigenvalue, 
i.e., the energy eigenvalues are zero or pure imaginary. 

Here we would like to summarize the property of flows in CSGT. 
As for the zero-energy flows we can stress the following three 
properties; they can be 
\hfil\break
(i) the absolutely energy-saving mechanism, 
\hfil\break
(ii) the mechanism including an enormous topological variety 
in terms of vortex patterns, and also 
\hfil\break
(iii) the perfect mechanism to recover the 
initial flow patterns in any disturbance by 
pouring arbitrary decaying flows with complex-energy eigenvalues. 
\hfil\break
The role of the flows with complex energies will be 
understood as short excitement mechanisms of the vortex patterns. 
We still have a lot of problems to overcome the present situation, but 
we may expect that 
the study of the zero- and complex-energy solutions in CSGT will open 
a new site in physics.

\hfil\break
{\bf 5. Remarks}

We have shown concrete examples of different types of 
vortex patterns accompanied by 
creations and annihilations of vortices by using only some low degree 
solutions of the 2D-PPB. 
We can, of course, present more complicated patterns 
by introducing the higher degree solutions, but the examples 
presented in the sections 3 
will be enough to show the fact that various vortex patterns 
can be reproducible in terms of the eigenfunctions of the 2D-PPB. 
As already noted that the zero-energy solutions in the 2D-PPB 
can be transformed to those in the potentials $V_a(\rho )$ by the 
conformal transformations [24,25], 
the stable patterns given in section 3.2 
can be transformed into the stable patterns of arbitrary potentials. 
This fact means that, as far as the stable patterns are concerned, 
there is an exact one-to-one correspondence between the patterns of 
the PPB and those of the other potentials. 
As for the time-dependent patterns we cannot present any concrete examples 
except the case of the 2D-PPB at this moment, 
but we may expect that similar vortex patterns as those given in section 
3.3 for the PPB will appear in other potentials, 
since all energy eigenstates with complex eigenvalues degenerate infinitely 
in all the potentials $V_a(\rho )$ as same as in the 2D-PPB. 
Anyhow we cannot exactly say about the problem before we find any solutions 
with complex eigenvalues in the other cases. 

Here we would like to note 2D-PPB. 
We do not know any physical phenomena that are described by 2D-PPB. 
We can, however, expect that most of weak repulsive forces in matters composed of 
many constituents will be approximated by PPBs as most of weak attractive forces 
are well approximated by harmonic oscillators. 
In general flows that go round a smooth hill of potential feel a weak 
repulsive force represented by a PPB [34-36]. 
Actually we see that when a charged particle is put in an infinitely 
long tube where same charged particles are uniformly distributed, 
the charged particle feels 2D-PPB. 
In non-neutral plasma  electrons being near the center will possibly be in 
a similar situation. 
In the plasma electro-magnetic interactions must be introduced. 
It should be noted that in the case of a charged particle in a magnetic field 
the vortex quantization given by (6) can be read as 
\begin{align}
 m \Gammav &=\oint _c (\nabla S -q{\bf A})\cdot d{\bf s} \nonumber \\
        &=\oint _c {\bf p}\cdot d{\bf s} - q\Phi ,
 \end{align} 
 where $q$ is the charge of the particle and 
 $\Phi$ is the magnetic flux passing through the enclosed surface. 
Analyzing vortex phenomena of non-neutral plasma in terms of the  
eigenfunctions of the 2D-PPB will be an interesting application.

The infinite freedom arising from the infinite degeneracy 
of the zero-energy solutions should be noticed. 
Such a freedom has never appeared in the statistical mechanics describing 
thermal equilibrium. 
The freedom is different from that generating the usual entropy and then 
temperatures, because the freedom does not change real energy 
observed in experiments at all. 
A model of statistical mechanics for the new freedom has been proposed 
and some simple applications have been performed in the case of 
1D-PBB [37-39]. 
The model is applicable to slowly changing phenomena in the time 
evolutions, because the PPB has only pure imaginary energy eigenvalues 
in the 1-dimension. 
In the present model of the 2-dimensions, however, 
we have the infinite degree of freedom 
arising from the zero-energy solutions that have no time evolution. 
The huge degeneracy of the zero-energy solutions can provide the huge 
variety in every energy eigenstate, which will be identified by 
the vortex patterns. 
In such a consideration the vortex patterns will be understood as 
the topological properties of flows. 
How this freedom should be counted in statistical mechanics is an important 
problem in future considerations. 

Finally we would like to comment on the meaning of the eigenstates in CSGT. 
As already noted, the eigenfunctions in CSGT are generally not normalizable, 
and then the probability and the probability current cannot have a definite 
meaning. 
This fact means that the probabilistic interpretation for the eigenfunctions 
cannot be introduced in CSGT. 
How should we interpret the eigenfunctions in CSGT? 
From the discussions presented in this paper we find out a possible idea that 
the quantization in terms of Gel'fand triplets describes 
the quantization of flows. 
Flows are, of course, composed of many particles, and then the probability 
used in the description of one particle motions cannot be introduced. 
The magnitudes of the eigenfunctions should be considered to be proportional to 
the densities of the flows like the intensity of beams in scattering processes. 
Thus the normalizations of wavefunctions 
expessed by the linear combinations of the eigenfunctions lose the meaning in CSGT. 
We can, however, fond out that the energies are quantized 
as discrete or continuous numbers including complex numbers, 
and also the flows expressed by the eigenfunctions 
interfere. 
Though quantities in CSGT are 
in general not directly observed except eigenvalues such as energies, 
we see that velocities have a special role that they are observables 
being definable only on CSGT. 
Vortex patterns that are determined only from nodal points of wavefunctions 
are also good observables to investigate solutions 
in CSGT. 
As shown in section 3, 
we can actually see the interferences among flows 
through the investigation of the vortex patterns. 
We may say that hydrodynamical approach will be an interesting trial 
to investigate physics in CSGT.

\pagebreak

\pagebreak

  \begin{figure}
   \begin{center}
    \begin{picture}(200,200)
     \thicklines
     \put(0,100){\vector(1,0){200}}
     \put(100,0){\vector(0,1){200}}
     \put(90,88){$0$}
     \put(205,98){$x$}
     \put(98,205){$y$}
     
      \put(150,150){$\bullet $}
     \put(150,50){$\bullet $}
     \put(48,150){$\bullet $}
     \put(48,50){$\bullet $}

     \put(151,98){$\cdot $}
     \put(48.5,98){$\cdot $}
     \put(156,105){$|c|/\beta $}
     \put(14,105){$-|c|/\beta $}
     \put(98,150){$\cdot $}
     \put(98,50){$\cdot $}
     \put(106,150){$|c|/\beta $}
     \put(103,50){$-|c|/\beta $}

     \put(152.4,150){\vector(0,-1){47}}
     \put(152.5,52){\vector(0,1){46}}
     \put(50.4,52){\vector(0,1){46}}
     \put(50.4,150){\vector(0,-1){47}}
    \end{picture}
   \end{center}
   \caption[]{Two pairs annihilation pattern.}
   \label{fig:1.7}
  \end{figure}
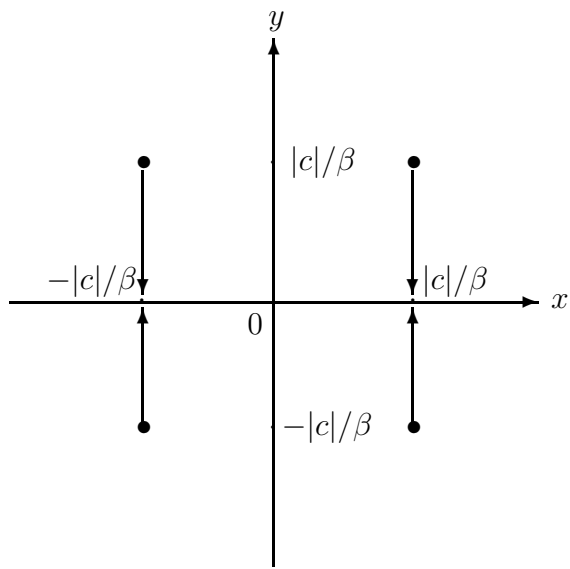

       \begin{figure}
    \begin{center}
    \begin{picture}(200,200)
     \thicklines
     \put(0,100){\vector(1,0){200}}
     \put(100,0){\vector(0,1){200}}
     \put(90,88){$0$}
     \put(205,98){$x$}
     \put(98,205){$y$}

     \put(183,159){$\bullet$}
     \put(15,40){$\bullet $}
     \put(185.9,161.5){\vector(-3,-2){7}}
     \put(17.5,43.3){\vector(3,2){7}}
     \put(153,139){$\bullet$}
     \put(45,60){$\bullet $}
     \put(156,141.8){\vector(3,2){7}}
     \put(48.2,62.4){\vector(-3,-2){7}}

     \put(15,159){$\bullet$}
     \put(153,60){$\bullet $}
     \put(17.9,161.7){\vector(3,-2){7}}
     \put(155.5,62.3){\vector(3,-2){7}}
     \put(45,139){$\bullet$}
     \put(183,40){$\bullet $}
     \put(48.1,142.1){\vector(-3,2){7}}
     \put(185.9,43.0){\vector(-3,2){7}}
     
    \end{picture}
   \end{center}
   \caption[]{Four pairs creation and annihilation. 
        }
   \label{fig:1.8}
    \end{figure}
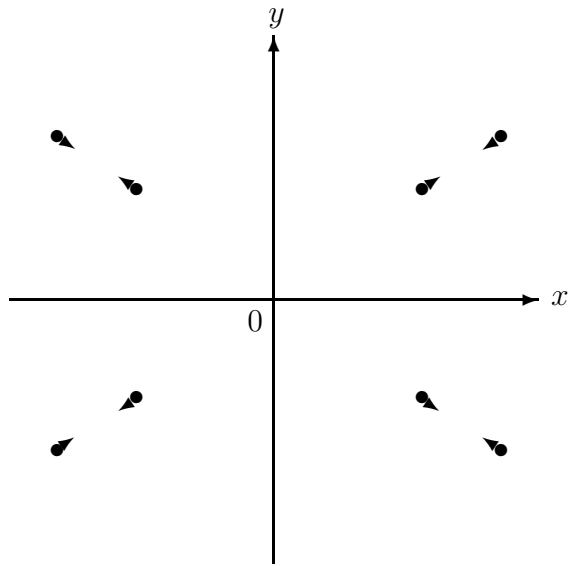

\end{document}